\documentstyle[11pt,aaspp4]{article}

\slugcomment{Submitted to The Astrophysical Journal}

\lefthead{}
\righthead{}

\begin{document}

\title{Three-Dimensional Magnetohydrodynamic Simulations of Spherical Accretion}

\author{Igor V. Igumenshchev\altaffilmark{1,2} and Ramesh Narayan}
\affil{Harvard-Smithsonian Center for Astrophysics, 60 Garden Street,
Cambridge MA 02138}

\altaffiltext{1}{Laboratory for Laser Energetics, University of Rochester,
250 East River Road, Rochester NY 14623-1299.}

\altaffiltext{2}{Institute of Astronomy,
48 Pyatnitskaya Ulitsa, 109017 Moscow, Russia.}

\begin{abstract}

We present three-dimensional numerical magnetohydrodynamic simulations
of radiatively inefficient spherical accretion onto a black hole. The
simulations are initialized with a Bondi flow, and with a weak,
dynamically unimportant, large-scale magnetic field.  The magnetic
field is amplified as the gas flows in.  When the magnetic pressure
approaches equipartition with the gas pressure, the field begins to
reconnect and the gas is heated up.  The heated gas is buoyant and
moves outward, causing line stretching of the frozen-in magnetic
field.  This leads to further reconnection, and more heating and
buoyancy-induced motions, so that the flow makes a transition to a
state of self-sustained convection.  The radial structure of the flow
changes dramatically from its initial Bondi profile, and the mass
accretion rate onto the black hole decreases significantly.

Motivated by the numerical results, we develop a simplified analytical
model of a radiatively inefficient spherical flow in which convective
transport of energy to large radii plays an important role.  In this
``convection-dominated Bondi flow'' the accretion velocity is highly
subsonic and the density varies with radius as $\rho\propto R^{-1/2}$
rather than the standard Bondi scaling $\rho\propto R^{-3/2}$.  We
estimate that the mass accretion rate onto the black hole
correspondingly scales as $\dot M \sim (R_{in}/R_a)\dot M_{Bondi}$,
where $R_{in}$ is a small multiple of the Schwarzschild radius of the
black hole and $R_a$ is the ``accretion radius'' at which the ambient
gas in the surrounding medium is gravitationally captured by the black
hole.  Since the factor $R_{in}/R_a$ is typically very small, $\dot M$
is significantly less than the Bondi accretion rate.
Convection-dominated Bondi flows may be relevant for understanding
many astrophysical phenomena, e.g. post-supernova fallback and
radiatively inefficient accretion onto supermassive black holes,
stellar-mass black holes and neutron stars.

\end{abstract}


\section{Introduction}

The classic problem of adiabatic spherical accretion onto a compact
gravitating mass has been studied by many authors.  The hydrodynamic
version of this problem was solved by Bondi (1952) who showed that
inside a certain capture radius, the radial velocity of the accreting
gas varies as $v_R\propto R^{-1/2}$, where $R$ is the radius.
Correspondingly, the density varies as $\rho\propto R^{-3/2}$.

If the gas in a Bondi flow has a frozen-in magnetic field, the field
lines are stretched in the radial direction and compressed in the
transverse direction, so that the radial component of the field is
amplified according as $B_R\propto R^{-2}$.  The magnetic energy
density then varies as $\epsilon_m=B^2/8\pi\propto R^{-4}$.  However,
the gravitational energy density of the gas varies only as
$\epsilon_{grav}=\rho GM/R\propto R^{-5/2}$. Thus, for a sufficiently
small radius, one formally has $\epsilon_m\gg\epsilon_{grav}$, which
is physically inconsistent since the magnetic energy ultimately is
derived from the gravitational binding energy of the accreting gas.

Shvartsman (1971) proposed that the conversion of gravitational energy
into magnetic energy is accompanied by turbulence which tangles the
magnetic field lines.  He suggested that, as the magnetic and
gravitational energies approach equipartition, field reconnection
ensures that $\epsilon_m$ does not exceed $\epsilon_{grav}$.
Reconnection will be accompanied by dissipation (Bisnovatyi-Kogan \&
Ruzmaikin 1974; M\'esz\'aros 1975) which will heat the gas.  As a
result, the thermal energy of the gas is also likely to come into
rough equipartition with the other energies.  These modifications have
been recognized for many years and have been incorporated into models
of spherical accretion flows.

To our knowledge, all published studies of the magnetized spherical
accretion problem have assumed that the magnetic field causes no
serious {\it dynamical} effects on the flow.  In particular, all
authors assume that the velocity and density scale exactly as in the
Bondi solution, namely $v_R\propto R^{-1/2}$, $\rho\propto R^{-3/2}$.
(The following is an incomplete list of relevant papers on the
subject: Zeldovich \& Novikov 1971; Shapiro 1973a,b; Ipser \& Price
1977, 1982, 1983; Maraschi et al. 1979; Maraschi, Roasio, \& Treves
1982; Scharlemann 1981, 1983; Treves, Maraschi, \& Abramowicz 1988;
Turolla \& Nobili 1989; Nobili, Turolla, \& Zampieri 1991; Mason \&
Turolla 1992; Mason 1992; Melia 1992; Kowalenko \& Melia 1999; Coker
\& Melia 2000; many other papers discuss spherical accretion without
explicitly considering magnetic fields and heating, e.g. Ostriker et
al. 1976; Park \& Ostriker 1989; Houck \& Chevalier 1991; Zampieri et
al. 1998.)

In this paper we show that the magnetic field can play an important,
perhaps even dominant, role in the dynamics of spherical accretion
flows.  The influence is both direct, through the action of the
electromotive force, and indirect, through the entropy generated in
the process of field reconnection.  The latter leads to convection,
which drastically changes the flow structure relative to the Bondi
solution.  In particular, the density profile becomes much less
centrally peaked than the Bondi $R^{-3/2}$ profile, and the mass
accretion rate is reduced significantly below the Bondi rate.
Interestingly, the flow resembles the recently discovered
hydrodynamical convection-dominated accretion flow solution (CDAF,
Narayan, Igumenshchev, \& Abramowicz 2000; Quataert \& Gruzinov 2000;
Stone, Pringle, \& Begelman 1999; Igumenshchev \& Abramowicz 2000,
2001; Igumenshchev, Abramowicz, \& Narayan 2000; Ball, Narayan, \&
Quataert 2001).

The rest of the paper is organized as follows. In \S2 we describe our
numerical method, and the details of our initial and boundary
conditions. In \S3 we present results of numerical simulations, and in
\S4 we describe an approximate self-similar solution which includes
the effects of plasma heating and convection. In \S5 we discuss some
implications of the results.

\section{Simulation Technique}

\subsection{Magnetohydrodynamic Equations}

We solve the equations of resistive MHD in the one-fluid
approximation,
\begin{equation}
{d\rho\over dt} + \rho{\bf\nabla\cdot v} = 0,
\end{equation}
\begin{equation}
\rho{d{\bf v}\over dt} = -{\bf\nabla} (P+Q) - \rho{\bf\nabla}\Phi +
{1\over 4\pi}({\bf\nabla}\times{\bf B})\times {\bf B},
\end{equation}
\begin{equation}
\rho{d\epsilon\over dt} = -(P+Q){\bf\nabla\cdot v} + {1\over
4\pi}\eta{\bf J}^2,
\end{equation}
\begin{equation}
{\partial{\bf B}\over \partial t} = {\bf\nabla}\times({\bf v}\times{\bf B} -
\eta{\bf J}),
\end{equation}
where $\rho$ is the density, ${\bf v}$ is the velocity, $P$ is the
pressure, $\Phi$ is the gravitational potential, ${\bf B}$ is the
magnetic induction, $\epsilon$ is the specific internal energy, ${\bf
J}={\bf\nabla}\times{\bf B}$ is the current density, and $\eta$ is the
resistivity.  The terms involving $Q$ in equations (2) and (3)
correspond to an artificial viscosity which is introduced to correctly
treat shocks.  We adopt the ideal gas equation of state,
\begin{equation}
P=(\gamma-1)\rho\epsilon,
\end{equation}
with an adiabatic index $\gamma=5/3$.  We assume that there is no
radiative cooling.

We take the compact mass at the center to be a black hole of mass $M$
and we use a pseudo-Newtonian gravitational potential (Paczy\'nski \&
Wiita 1980) to mimic the effects of general relativity,
\begin{equation}
\Phi=-{GM\over R-R_g}, \qquad R_g={2\,GM\over c^2},
\end{equation}
where $R_g$ is the gravitational radius.

\subsection{Numerical Method}

We numerically solve equations (1)--(4) by using an extension of the
original PPM algorithm developed by Colella \& Woodward (1984) for
hydrodynamics.  We use the Lagrangian version of the PPM algorithm
with operator splitting.  The Riemann solver takes into account the
non-linear interaction of the fast MHD waves when calculating
time-updates of the density, velocity and internal energy in equations
(1)--(3).  The components of the Lorentz force in the equation of
motion (2) are calculated using a solution of the Lagrangian
characteristic equations for Alfv\'en waves, as in the method of
characteristics (MOC) of Stone \& Norman (1992). We use dimension
splitting when solving the equations in three dimensions.

We replace the induction equation (4) by an equivalent equation for
the vector potential ${\bf A}$,
\begin{equation}
{\partial{\bf A}\over\partial t} = {\bf v}\times{\bf B} -
\eta{\bf J},
\end{equation}
where ${\bf B}={\bf\nabla}\times{\bf A}$.  This approach guarantees
that the constraint ${\bf\nabla\cdot B}=0$ is satisfied to within grid
approximation errors at each time $t$. The first term on the right
hand side of equation (7) is calculated using a modified version of
the Eulerian MOC algorithm (Stone \& Norman 1992).  (The modification
is due to our use of a different representation of the location of the
Alfv\'en characteristics domain for calculating the averaged values of
the components of ${\bf v}$ and ${\bf B}$.)

The code operates on a three-dimensional Cartesian grid.  In order to
adequately resolve the large dynamic range of spatial scales spanned
by astrophysical accretion flows, we employ a nested grid (see Fig. 1)
similar to that employed by Igumenshchev et al. (2000) for simulating
hydrodynamical CDAFs in three dimensions.  In the present
calculations, we have used five subgrids with $64\times 64\times 64$
cells in each subgrid.  The cell size in the innermost subgrid is
$\Delta_1=0.5 R_g$.  Each succeeding subgrid has its cell size
increased by a factor of 2, and so the outermost subgrid has
$\Delta_5=8 R_g$, and covers a cube of size $256 R_g \times 256 R_g
\times 256 R_g$.  In practice, we used only a quarter of the full
cubic domain, by focusing on a 90 degree wedge around the $z$ axis.
Thus, we employed $32\times32\times64$ cells along $xyz$, and used
periodic boundary conditions in the azimuthal direction.

\begin{figure}
\plotone{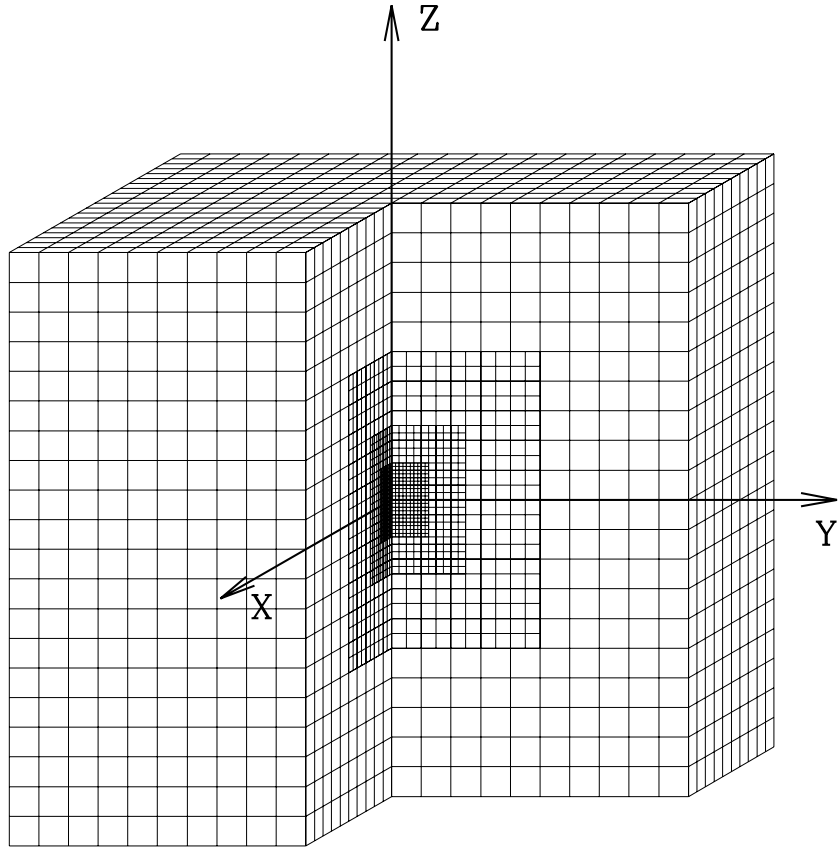}
\caption{Schematic view of the nested Cartesian grid used in the
simulations.  The example shown has four sub-grids, each of size
$20\times 20\times 20$.  In the actual calculations five sub-grids
were used with dimensions $64\times 64\times 64$. The black hole is
located at the origin.
\label{fig1}}
\end{figure}

\placefigure{fig1}

We define the quantities $\rho$, $\rho{\bf v}$ and $\rho\epsilon$ at
the centers of cubic cells, and the components of ${\bf B}$ at the
corresponding cell edges.  By using continuous piece-wise parabolic
approximations of ${\bf v}$ and ${\bf B}$ when solving equation (7),
the code avoids the ``explosive'' instability which is found in MHD
codes based on the original MOC algorithm (Clarke 1996).

The energy equation in numerical MHD needs to be handled with care.
Because of the finite spatial resolution, field lines can reconnect,
leading to a loss of magnetic energy without a compensating increase
in the internal energy of the gas.  This leads to uncontrolled energy
loss in the simulation (e.g. Stone \& Pringle 2001; Igumenshchev \&
Abramowicz 2001).

To fix the problem, we introduce an explicit artificial resistivity
$\eta$ and set its magnitude to be larger than the effective numerical
resistivity associated with numerical reconnection.  Following Stone
\& Pringle (2001), we choose
\begin{equation}
\eta=\eta_0{|{\bf\nabla}\times{\bf B}|\over\sqrt{4\pi\rho}}\Delta^2,
\end{equation}
where $\eta_0$ is a dimensionless parameter and $\Delta$ is the grid
spacing.  The magnetic Reynolds number corresponding to this
resistivity is
\begin{equation}
Re_m\simeq {1\over\eta_0}\left({L\over\Delta}\right)^2,
\end{equation}
where $L$ is the characteristic spatial scale of the problem.

Ryu, Jones, \& Frank (1995) have shown that the effective Reynolds
number due to numerical resistivity has the same functional dependence
on $L$ and $\Delta$ as in equation (9), and they estimate the
corresponding coefficient $(\eta_0)_n$ to be in the range $0.2-0.5$
for their second-order MHD code based on the total variation
diminishing scheme.  We expect that our code has the same or even
lower numerical resistivity.  For this reason, we have used $\eta_0$
in the range of $0.3-0.5$.  Test runs show a significant improvement
in energy conservation when we choose $\eta_0$ in this range.

In principle, the non-uniform nested numerical grid (Fig.~1) could
introduce perturbations in the flow at the interfaces between subgrids
because of differences in the numerical viscosity and resistivity on
the two sides of the boundary.  We checked for this effect in rotating
hydrodynamical accretion flows (Igumenshchev et al.  2000), and found
that the perturbations were small and had no significant effect on the
global flow.  The present simulations of MHD accretion flows again
show only minor effects at sub-grid interfaces.

\subsection{Initial and boundary conditions}

We initialize our simulations with a spherically symmetric flow,
described by the following self-similar Bondi solution for a
$\gamma=5/3$ gas in the Newtonian gravitational potential of a point
mass $M$,
\begin{equation}
\rho(R)={\dot{M}\over 4\pi}{R^{-3/2}\over v_0\sqrt{GM}}, \quad
v_R(R)=-v_0\sqrt{GM\over R}, \quad
\epsilon(R)={3\over 5}\left(1-{v_0^2\over 2}\right){GM\over R},
\end{equation}
where $\dot{M}$ is the mass accretion rate and $v_0$ is a dimensionless
parameter which can take any value in the range
$0\le v_0\le\sqrt{2}$.  In all our simulations we have taken
$v_0=1$, which corresponds to supersonic Bondi accretion with
Mach number ${\cal M}=1.7$.  We take the initial magnetic field to be
uniform with only one non-zero component: $B_z$.  We specify the initial
strength of the magnetic field $B_z$ by a parameter $b_0$ defined as follows,
\begin{equation}
{B_z^2\over8\pi}=b_0^2{GM\rho_{out}\over R_{out}},
\end{equation}
where $\rho_{out}$ is the density of the gas at the outer radius of
the grid $R_{out}=256\,R_g$.  We choose $b_0^2\ll1$, which ensures
that the magnetic field has negligible influence on the flow dynamics
early in the simulation.

At the outer boundary we assume that, at all times, the density and
the velocity are given by the Bondi solution (10) and the magnetic
field is equal to its initial value.  At the inner boundary, close to
the black hole horizon, we assume absorbing conditions.  Specifically,
any matter that crosses the inner radius $R_{in}=2 R_g$ is extracted
from the computational domain, and the magnetic terms in the equation
of motion (2) are switched off.

\section{Numerical Results}

\placetable{tbl-1}

We have calculated four models, A, B, C, D, with the parameters listed
in Table~1.  All the models begin with a weak magnetic field
($b_0^2\ll1$), which has a negligible dynamical effect on the flow.
Therefore, the spherical Bondi solution with which the flow is
initialized is stable and persists for some time.  As the gas flows
in, the strength of the magnetic field increases, with the most rapid
increase occurring in the innermost region.  At a certain critical
time the field becomes strong enough to modify the flow dynamics.

Figure~2 shows the evolution of the mass accretion rate onto the black
hole for Models~A (thin line), B (dashed line) and D (thick line).  At
the beginning of the simulations the accretion rates in all three
models experience a quick relaxation.  This initial relaxation is due
to the fact that there is a minor inconsistency between the
self-similar solution (10) and the absorbing inner boundary conditions
as well as the pseudo-Newtonian potential (6).  At the end of the
initial relaxation, the flow takes up a slightly modified steady state
configuration, and the mass accretion rate onto the black hole becomes
equal to the rate of mass input at the outer boundary.  At this point,
the flow is essentially the standard Bondi flow.  Model~D, which is a
pure hydrodynamic simulation, does not change any further after this
initial relaxation.  Models A, B and C, however, which have magnetic
fields, undergo significant evolution.

\placefigure{fig2}

Figure~3 shows the configuration of magnetic field lines in Model~A at
time $t=0.5$; we measure time in units of the free-fall time from the
outer radius $R_{out}$ of the computational domain.  We see
that the initially parallel field lines are pulled in towards the
black hole as they are swept in by the converging flow streamlines.
Note that, near the equatorial plane, oppositely directed magnetic
lines closely approach each other.  This leads to efficient
reconnection soon after this time.

\placefigure{fig3}

At $t \simeq 0.5$ in Model~A the strength of the magnetic field in the
vicinity of the black hole reaches equipartition with the gas
pressure, and the accretion rate experiences a sudden large drop.
This drop represents the effect of the amplified magnetic field, which
suppresses accretion in the equatorial zone and forces matter to
accumulate within a ``core'' region in the vicinity of the black hole.
Because the initial Bondi flow is supersonic, an MHD shock forms where
the inflowing matter meets the core.  The post-shock gas is sub-sonic
and sub-Alfvenic.  In Model~B the same drop in accretion rate is
again seen, except that it happens at a later time ($t\approx 1$).
This model starts with a weaker field and therefore it takes longer
for the magnetic pressure to build up to equipartition strength.
Model~C, with a smaller value of the resistivity parameter $\eta_0$,
has an evolution almost identical to that of Model~A.  All three
models relax to a new state in which the mass accretion rate is
several times less than the mass supply rate.  The accretion rates in
Models~A, B and C continue to evolve slowly with time and by the end
of the simulations ($t\sim8$), the rates are roughly an order of
magnitude less than the mass supply rate.  The accretion rates are
highly variable, however, reflecting the unstable nature of the flows.
All three models have very similar final states.

Here we discuss Model~A in detail as a representative example.  As
already mentioned, a dense core is formed in the innermost region of
the accretion flow once the magnetic field reaches equipartition with
the gas pressure.  The core is bounded by a quasi-spherical shock.
The size of the core and the mass contained in it increase with time
due to the accumulation of matter; matter flows onto the core from the
outside at a rate determined by the rate of mass input at the outer
boundary, while mass flows out of the core into the black hole at a
much lower rate (roughly a factor of ten lower).

Figures~4, 5 and 6 show the distribution of density, magnetic field
lines and velocity streamlines in the core region at a relatively late
time, $t=7.88$.  The influence of the large scale magnetic field
introduced through the outer boundary condition is still visible in
the polar direction.  The magnetic field here is almost radial and has
its maximum strength.  Correspondingly, the density takes its minimum
value, so that the plasma $\beta\equiv P_g/P_m$ is quite small,
$\beta\sim 0.1$.  The accreting matter moves along the magnetic lines
with supersonic velocity; the velocity is roughly of order the Alfven
speed.  Because of the high velocity there is no accumulation of
matter.

\placefigure{fig4}
\placefigure{fig5}
\placefigure{fig6}

As Figures~4--6 show, the picture is very different in the equatorial
region (``equator'' and ``pole'' are defined with respect to the
large-scale magnetic field).  The topology of the magnetic field is
much more tangled and the gas is denser.  The accumulated matter is
highly inhomogeneous, with filaments of higher and lower density being
sandwiched between each other (Fig.~4).  The inhomogeneous structure
is the result of convection in the core driven by the heat released
during episodes of magnetic reconnection.

The first reconnection event occurs at a time slightly later than that
shown in Figure~3.  The reconnection and corresponding energy release
happens exactly at the equatorial plane, in the vicinity of the inner
boundary, where the magnetic field is strongest and where oppositely
directed field lines approach each other most closely.  After the
reconnection event the local strength of the magnetic field is reduced
to a sub-equipartition level (as visualized by Shvartsman 1971) and
the gas is heated by the energy released in the reconnection.  The
heated gas expands and forms a convective blob which moves outward as
a result of buoyancy force.  During the motion of the hot blob through
the ambient medium it deforms magnetic lines and causes new
reconnection events.  This leads to the formation of other hot blobs
which again become convective.  As a result, the convective motions
are self-sustained and lead to turbulence in the core.  The magnetic
field in the convection zone is on average close to equipartition
strength, with $\beta\simeq 1$--$10$.

Figure~5 shows the tangled magnetic field configuration after the
convection has become fully developed, and Fig. 6 shows the chaotic
velocity streamlines.  We see numerous vortices and circulation
patterns.  The most efficient convection occurs at intermediate angles
between the poles and the equator.  Somewhat less efficient convective
motions are present in the equatorial plane, and there is no
convection in the polar regions.

\section{Self-Similar Solutions}

To better understand the physics of radiatively inefficient magnetized
spherical accretion, we consider here a steady radial flow and seek
self-similar solutions of the equations.  By the assumption of
self-similarity, we expect the plasma $\beta$ to be independent of
radius $R$.  Hence we write the gas, magnetic and total pressure as
\begin{equation}
P_g\equiv\rho c_s^2, \qquad
P_m={1\over\beta}\rho c_s^2, \qquad
P_{tot}=P_g+P_m={\beta+1\over\beta}\rho c_s^2,
\end{equation}
where $c_s$ is the isothermal sound speed of the gas.  The radial
momentum equation and the gas energy equation take the form
\begin{equation}
{d\over dR}\left({v_R^2\over 2}\right)
=-{1\over\rho}{dP_{tot}\over dR}-{GM\over R^2},
\end{equation}
\begin{equation}
\rho v_R T{ds \over dR}\equiv
\rho v_R \left[ {1\over\gamma-1}{dc_s^2\over dR}-{c_s^2\over\rho}
{d\rho\over dR}\right]
=-{1\over R^2}{d\over dR}(R^2 F_{c})+Q_{diss},
\end{equation}
where $v_R$ is the radial velocity (assumed negative for accretion),
$s$ is the specific entropy of the gas, $\gamma$ is the adiabatic
index, and $F_c$ is the outward flux of energy due to convection.  The
term on the left hand side of equation (14) describes the inward
advection of energy, the first term on the right hand side is the
divergence of the convective energy flux, and $Q_{diss}$ is the rate
of heating of the gas by dissipation.

Following Narayan \& Yi (1994, see also Kato, Fukue, \& Mineshige 1998;
Narayan et al. 2000), we use a simple parametric form to represent the
convective flux,
\begin{equation}
F_{c}=-\alpha_{c}c_s R \rho T{ds\over dR},
\end{equation}
where $\alpha_c$ is a dimensionless constant.  For the heating term,
$Q_{diss}$, we note that there are at least two sources of
dissipation: (i) energy release through magnetic reconnection, and
(ii) viscous and resistive dissipation at small scales as a result of
a turbulent cascade.  It is not possible to model these processes in
detail.  Instead, we note that the ultimate source of energy is the
gravitational potential energy of the accreting gas, and so we write
\begin{equation}
Q_{diss}=-\alpha_d{v_R\over R}\rho{GM\over R},
\end{equation}
where $\alpha_d$ is another dimensionless constant (the negative sign
is because $v_R<0$).

We consider self-similar flows in which the various variables behave
as power laws in radius (see the analogous discussion of CDAFs in
Narayan et al. 2000),
\[c_s(R)=c_0 v_K\propto R^{-1/2},\]
\begin{equation}
\rho(R)= \rho_0 R^{-a},
\end{equation}
\[v_R(R)= {\dot{M}\over4\pi R^2\rho}=-v_0 v_K\left(R_g\over R\right)^{(3/2-a)}
\propto R^{-2+a},\] where $v_K=\sqrt{GM/R}$ is the Keplerian velocity,
and $a$ is a power-law index which takes one of two values, 3/2 or 1/2
(see Narayan et al. 2000 and Quataert \& Gruzinov 2000).  The
coefficients $c_0$ and $v_0$ are dimensionless coefficients whose
values are determined by substituting the solution into the two
conservation equations (13) and (14).  The coefficient $\rho_0$ is
proportional to the accretion rate $\dot M$ and scales out of the
problem.

The simplest case to consider is the pure hydrodynamic problem, in
which gas with $\gamma=5/3$ accretes onto a black hole, with neither
convection nor dissipative heating.  The self-similar solution with
$a=3/2$ automatically satisfies the energy equation, while the
momentum equation gives
\begin{equation}
v_0^2=2-{5(\beta+1)\over\beta}c_0^2.
\end{equation}
We then have a family of solutions in which $v_0$ is a free parameter
and $c_0$ is determined by equation (18) (with $\beta\to \infty$ since
there is no magnetic field),
\begin{equation}
c_0^2={1\over5}(2-v_0^2).
\end{equation}
This is the self-similar solution given in equation (10), which was
used to initialize the numerical simulations.  The same self-similar
solution is often used even for magnetized spherical accretion, under
the assumption that the field will achieve equipartition and thereby
have a self-similar scaling.  But this is not correct.  When there is
a magnetic field, there is bound to be reconnection (see the
discussion in \S1) and this means that the entropy of the gas will
increase inward.  This is not consistent with the assumed scaling
($a=3/2$, $\gamma=5/3$, cf. Quataert \& Narayan 1999).

In the presence of magnetic fields, we need to consider the more
general problem with finite values of $\beta$, $\alpha_c$ and
$\alpha_d$, and a general value of $\gamma$.  Let us first assume that
$a$ takes the standard Bondi value of 3/2.  In this case, all the
terms in the energy equation (14) are of the same order (all scale as
$R^{-4}$), and the energy equation gives the following relation
between $v_0$ and $c_0$,
\begin{equation}
\left({1\over\gamma-1}-{3\over2}\right)(v_0-\alpha_c c_0)
c_0^2=\alpha_d v_0.
\end{equation}
This relation, combined with equation (18) from the momentum equation,
allows us to solve uniquely for $v_0$ and $c_0$ for given values of
$\beta$, $\alpha_c$, $\alpha_d$ and $\gamma$.  However, not all
combinations of these parameters lead to a physical solution.  For
instance, consider the case when $\alpha_c c_0 \ll v_0$.  Equation
(20) then gives
\begin{equation}
c_0^2 = {\alpha_d\over \left({1\over \gamma-1}-{3\over2}\right)},
\end{equation}
which shows that $\gamma$ has to be less than 5/3 if $c_0^2$ is to be
positive and finite.  In fact, the constraint on $\gamma$ is even more
severe.  This can be seen by substituting (21) in (18) and solving for
$v_0^2$.  The requirement that $v_0^2>0$ leads to
\begin{equation}
{1\over \gamma-1} > {3\over2} + {5(\beta+1)\alpha_d\over 2\beta},
\end{equation}
which gives an upper limit on $\gamma$ that is smaller than 5/3.

The reason for these constraints is that the value $\gamma=5/3$ is a
singular case for a self-similar solution with $a=3/2$.  This fact was
noted by Quataert \& Narayan (1999) for the case of a rotating viscous
flow, but the same argument applies here as well.  When the condition
$\alpha_c c_0 \ll v_0$ which was used to derive (21) is not satisfied,
the inequality (22) becomes modified to a more complicated relation.
However, the qualitative features remain the same.  Specifically, we
find that $\gamma$ has to be smaller than a certain value (which is
less than 5/3) in order to have a physical solution.

Let us next consider a self-similar solution with $a<3/2$.  Now, the
various terms in the energy equation are no longer of the same order.
The entropy advection term and the heating term still vary as
$R^{-4}$, but the term describing the divergence of the convective
flux varies as $R^{-5/2-a}$.  The latter term dominates at large $R$.
Since there is no other term to balance this term, the only way to
satisfy the energy equation is to ensure that its coefficient is equal
to zero.  This requires $a=1/2$.

Let us now set $a=1/2$.  As we have just argued, the energy equation
is automatically satisfied (to leading order) at large $R$.  The
momentum equation also becomes simpler since the term involving $v_R$
is smaller than the other two terms and may be neglected.  Thus we
find
\begin{equation}
c_0^2={2\beta\over3(\beta+1)}.
\end{equation}
The value of the parameter $v_0$ is not uniquely determined by this
analysis, but is fixed by boundary conditions.  In particular, close
to the black hole, where the various terms in the energy equation
become of comparable order (in fact, the convective term probably
becomes less important than the other two terms), the flow will make a
transition to a different regime; this is also the region where the
flow matches onto the absorbing boundary condition at the black hole.
We do not discuss the physics of this region as it is beyond the scope
of the present paper.

A feature of the $a=1/2$ self-similar solution is that $\gamma=5/3$ is
not a singular case (the solution is singular when $\gamma=3$, but
this has no practical consequences).  Thus, for an accretion flow with
$\gamma=5/3$, the $a=1/2$ solution appears to be more robust than the
$a=3/2$ solution.  For lower values of $\gamma$, both the $a=3/2$
solution and the $a=1/2$ flow may be allowed and it is not a priori
obvious which solution would be picked by nature.  We suspect that the
$a=1/2$ solution is the flow of choice under most conditions (see also
the discussion of the $a=1/2$ law by Gruzinov 2001).

The two solutions discussed above are very different from each other.
In the $a=3/2$ flow, convection is a relatively minor perturbation and
the energy balance is primarily between energy advection and
dissipative heating.  This flow is very similar to the standard Bondi
accretion flow, with only minor changes in the values of some
coefficients.  In contrast, in the new $a=1/2$ solution, the
convective flux dominates the energy equation.  We therefore refer to
it as a convection-dominated Bondi Flow (CDBF).  This flow deviates
remarkably from the standard Bondi solution; in fact, it resembles the
CDAF solution.

The CDBF has a steady outward flux of
energy due to convection.  The convective luminosity is
\begin{equation}
L_c=4\pi R^2F_c= {\alpha_c c_0^3\over 2v_0}
\left({1\over\gamma-1}-{1\over2}\right)\dot Mc^2
\equiv \varepsilon_c\dot Mc^2.
\end{equation}
Thus, convection transports a fraction of the binding energy of the
accreting gas outward.  The efficiency of this process, described by
the coefficient $\varepsilon_c$, depends on details of the flow which are
not easy to determine from first principles.  In the case of a CDAF,
where again there is an analogous relation, numerical simulations by
Igumenshchev \& Abramowicz (2000; see also Narayan et al. 2000;
Igumenshchev et al. 2000) give $\varepsilon_c\sim 0.001-0.01$.  It is
likely that a CDBF also has a similar efficiency.

\section{Summary and Discussion}

The main result of this paper is that radiatively inefficient
spherical (i.e. non-rotating) accretion of magnetized plasma onto a
compact mass has very different properties compared to spherical
accretion of unmagnetized gas.  The unmagnetized problem is described
by pure hydrodynamics.  It was solved by Bondi (1952) and has been
widely applied.

Our results on the magnetized problem are based on three-dimensional
numerical MHD simulations.  The simulations are initialized with an
analytical self-similar Bondi flow (eq. [10]) and have initially a
weak, dynamically unimportant magnetic field ($b_0^2\ll1$, cf
eq. [11]).  The inner boundary conditions correspond to a black hole.
We find that the Bondi solution survives for a short period of time
with relatively minor changes.  However, during this time the magnetic
field becomes progressively stronger --- because of radial stretching
and transverse compression of the frozen-in field lines as the gas
flows in --- until the magnetic pressure builds up roughly to
equipartition with the gas pressure (Figure~3).

After this time, the structure of the flow changes dramatically.  The
equipartition magnetic field exerts a back-reaction on the
free-falling gas and slows it down in the ``equatorial'' region.  A
shock forms and gas accumulates in a ``core.''  The magnetic field
begins to reconnect to maintain the magnetic pressure slightly below
equipartition (plasma $\beta\sim1-10$).  The reconnection heats the
gas locally, and the resulting high entropy material moves outward
under buoyancy.  The outward motion causes further stretching and
amplification of the frozen-in field lines, leading to further
episodes of reconnection.  Before long, the plasma in the vicinity of
the black hole experiences fully developed turbulent convection.  With
time, the convective core grows in size, as more magnetized material
from the outside is added to it, while mass flows into the black hole
at a much smaller rate.  The flow in the convective core bears almost
no resemblance to the Bondi solution (see Figures~4, 5 and 6).  One
result of all this is that the accretion rate onto the black hole is
reduced significantly from the Bondi rate (Figure~2).  The flow does
not have a wind or outflow, as in the models of Blandford \& Begelman
(1999) and Das (2000), but this is not surprising since those authors
include additional ingredients such as angular momentum and viscosity
(Blandford \& Begelman, see also Narayan \& Yi 1994) and pair physics
(Das).

The principal physical effects operating in our numerical simulations
may be understood by considering the energetics of the accreting
magnetized plasma.  We can identify three major stages in which the
gravitational binding energy of the accreting gas is transformed into
other forms of energy:
\[ {\rm (gravitational~energy)}\]
\[   \downarrow \]
\[ {\rm (magnetic~energy)} \]
\[   \downarrow \]
\[ {\rm (thermal~energy)} \]
\[   \downarrow \]
\[ {\rm (convective~turbulent~energy,~leading~to~convective~energy~transport)}
\]
The first and second transformations, namely
(gravitational~energy)$\rightarrow$(magnetic~energy) and
(magnetic~energy)$\rightarrow$(thermal~energy) are well-known and have
been widely discussed (Shvartsman 1971; Bisnovatyi-Kogan \& Ruzmaikin
1974; M\'esz\'aros 1975; and numerous later papers, see \S1).  These
two processes build up the magnetic field to near-equipartition
strength and modify the thermal state of the gas (and its radiative
properties) relative to the standard Bondi flow.  However, they have
little effect on the overall dynamics of the flow.

The third transformation,
(thermal~energy)$\rightarrow$(convective~turbulence), does have a very
important effect on the dynamics, and this appears to have been
overlooked in previous studies.  The importance of convection is not
because it creates turbulent kinetic energy (which is just another
form of pressure, like magnetic and thermal pressure), but because it
causes energy {\it transport}.  Convection efficiently transports
energy from the deep interior of the flow, where the bulk of the
gravitational energy is released, to the outer regions of the flow.
Away from the center, the convective flux dominates the physics and
thus has a significant effect on the structure of the flow.

In the MHD simulations presented here, the mass accretion rate onto
the black hole is reduced by a factor of about ten relative to a Bondi
flow with the same outer boundary conditions (of density and sound
speed).  The reduction occurs because of the formation of the
convective core.  The mean accretion velocity in the core is highly
subsonic, and much smaller than the radial velocity in an equivalent
Bondi flow.  Only very close to the inner absorbing boundary does the
accretion become supersonic, in contrast to a standard Bondi flow
which has supersonic infall over a wide range of radius.  The radial
density profile is also very different from the $R^{-3/2}$ profile
found in a Bondi flow.

We should caution that, in our models, we have assumed that the
external medium has a uniform magnetic field.  This leads to a large
departure from spherical symmetry in the accretion flow at late times.
For instance, Figures~4--6 show a bipolar structure in the magnetic
field, with two ``poles'' (oriented parallel to the external field)
along which there is preferential accretion.  The situation we have
simulated would be realized if the coherence scale of the field in the
external medium is larger than the accretion radius $R_a$ --- the
radius at which the gas from the external medium is captured by the
gravitational pull of the accreting mass.  However, one could
visualize other situations in which the field is tangled on small
scales in the external medium, so that the accreting gas has several
distinct magnetic loops.

We have tried numerical experiments on accretion flows with small
scale magnetic field of different configurations.  However, we lacked
sufficient numerical resolution for these experiments; the field
underwent resistive dissipation before it could reach equipartition
with the gas pressure.  This demonstrates the importance of having
adequate numerical resolution.  We estimate that to conduct any
believable experiments with a non-uniform external field we will need
to increase the numerical resolution by a factor of at least~2--3.

Despite the above cautionary comment, the basic physical ideas we have
presented should be valid whatever be the topology of the field;
namely, line stretching and field amplification leads to reconnection,
which leads to gas heating, which leads to convection.  This chain of
argument requires merely that the magnetic field should be frozen into
the gas and that the magnetic field should build up to equipartition
strength before the gas falls into the black hole.  Since the magnetic
pressure grows as $R^{-4}$ whereas the gas pressure varies only as
$R^{-5/2}$ (\S1), the latter condition should be easily satisfied in
most cases of interest.

Another cautionary comment is related to the fact that our simulations
have not reached steady state.  The convective core grows slowly and
has reached a size of only about $80R_g$ by the end of our
simulations.  This is still a factor of 3 smaller than the size of the
grid.  In a real accretion flow we imagine that the convective core
would grow until it extends beyond $R_a$.  The convective flux would
then flow out into the external medium, and perhaps modify the medium
in the vicinity of $R_a$.  Ultimately, a steady state should result,
but the present simulations have not been run long enough to determine
the nature of the steady state.

In \S4 we present a simplified set of equations to describe spherical
accretion of a magnetized plasma.  The equations include two critical
pieces of physics: (i) heating of the gas by reconnection (and other
dissipative processes), and (ii) convective energy transport.
Depending on parameters, we find that there are two distinct
self-similar solutions of the equations.

One solution is not very different from the standard Bondi solution;
the density varies as $\rho\propto R^{-3/2}$ and the velocity varies
as $v_R\propto R^{-1/2}$.  This solution is possible whenever
convection is not very strong and when the adiabatic index $\gamma$ of
the gas is smaller than a limit which is less than the standard value
of 5/3 (cf. discussion below eqs [21, 22]).  For given boundary
conditions at the accretion radius $R_a$, the mass accretion rate in
this solution is similar to the Bondi accretion rate $\dot M_{Bondi}$.
Our numerical simulations, however, are not consistent with this
solution.  At this time we are not sure if the solution is relevant
for any radiatively inefficient spherical accretion flow with strong
fields.

The second solution is completely different from the Bondi solution.
It has density varying as $\rho\propto R^{-1/2}$ and velocity varying
as $v_R\propto R^{-3/2}$.  The scalings may be understood as follows.

The bulk of the energy generation in the accretion flow happens close
to the black hole.  Some fraction of this energy is transported
outward by convection.  At radii greater than a certain transition
radius $R_{in}$, whose value is uncertain but is probably no more than
a few tens of $R_g$, the convective luminosity $L_c$ completely
dominates over any local energy generation.  Thus, for $R>R_{in}$, we
expect $L_c$ to be practically independent of $R$; equivalently, the
convective flux $F_c$ must vary as $R^{-2}$.  Because the accretion
flow is assumed to be radiatively inefficient, the gas is virial and
has a sound speed $c_s\sim v_K$ (the Kepler or free-fall velocity).
Thus, there is only one velocity in the problem, namely $v_K\propto
R^{-1/2}$; therefore, the convective flux has to take the form
$F_c\sim\rho c_s^3 \sim\rho v_K^3 \sim\rho R^{-3/2}$.  Requiring the
convective flux to vary as $R^{-2}$ means that $\rho$ must scale as
$R^{-1/2}$.  Mass conservation, $\dot M=-4\pi R^2v_R\rho=$ constant,
then immediately gives $v_R\propto R^{-3/2}$.  Thus, the structure of
the flow is determined uniquely once we assume (i) that there is a
significant flux of energy outward due to convection, and (ii) that
there is no significant radiative cooling.

Because of the important role played by convection, we refer to this
kind of flow as a convection-dominated Bondi flow, or CDBF.  The
scalings sketched out above allow us to estimate the mass accretion
rate in this flow.  In the standard Bondi problem, where a mass $M$ is
embedded in a homogeneous medium of density $\rho_\infty$ and sound
speed $c_\infty$, the accretion radius is given by $R_a\sim
GM/c_\infty^2$; this is the radius at which the gravitational
free-fall velocity is equal to $c_\infty$.  The gas in the Bondi
solution flows in at essentially the free-fall velocity for $R\lesssim
R_a$; therefore, the mass accretion rate is given by $\dot
M_{Bondi}\sim4\pi R_a^2\rho_\infty c_\infty$.  In a CDBF, we expect
$v_R$ to be roughly equal to the free-fall velocity at $R\sim R_{in}$.
Since $v_R$ falls off as $R^{-3/2}$, this means that at $R\sim R_a$,
$v_R$ is smaller than the local free-fall velocity by a factor $\sim
R_{in}/R_a$.  We then estimate that
\begin{equation}
\dot M_{CDBF}\approx{R_{in}\over R_a}\dot M_{Bondi}.
\end{equation}
This shows that convection can have a profound effect on the mass
accretion rate in a magnetized spherical accretion flow.

The convective core region in the simulations presented in this paper
(Figures~4, 5 and 6) have features that are qualitatively similar to
the predictions of the self-similar CDBF solution.  Unfortunately, due
to limited spatial resolution, we have not been able to make
quantitative comparisons between the numerical results and the
predictions of the analytical model.  The averaged radial profiles of
density, velocity, gas and magnetic pressures show significant
oscillations, both in space and time.  They also show the influence of
the inner boundary (black hole) and outer boundary (the shock where
the free-falling gas meets the convective core), which cannot be
separated out because of the relatively small range of radius covered
by the simulations.  More extensive numerical work is required to
confirm the theoretical predictions in detail.  Also, in a real flow,
we expect the CDBF zone to extend all the way out to $R_a$, where it
would match the ambient density $\rho_\infty$ and ambient pressure
$\rho_\infty c_\infty^2$ of the external medium.  The simulations have
not reached steady state.  This is another reason why it is difficult
to compare the numerical results with the theoretical predictions.

We should note the close analogy between the CDBF solutions described
here and the viscous rotating (non-magnetic) CDAF solution (Narayan et
al. 2000; Quataert \& Gruzinov 2000).  The CDAF is more complex
because, in addition to the radial momentum equation and the energy
equation, it is also strongly influenced by the angular momentum
equation.  In particular, there is a competition between viscosity and
convection in the angular momentum balance, which plays an important
role in determining the structure of the solution.  Nevertheless, the
particular radial scalings seen in a CDAF, $\rho\propto R^{-1/2}$,
$v_R\propto R^{-3/2}$, are identical to those found in a CDBF, and
they result from the same physics identified above, namely the
presence of an energetically dominant convective flux and the absence
of radiative cooling.  Furthermore, the accretion rate in a CDAF is
reduced compared to that in an equivalent advection-dominated
accretion flow (ADAF, cf. Narayan \& Yi 1994; Abramowicz et al. 1995),
for the same outer boundary conditions, by a factor $\sim
R_{in}/R_{out}$, which is similar to the factor given in equation
(25).  The exact value of $R_{in}$ in the two problems depends on the
nature of the flow close to the black hole.  This is discussed in a
forthcoming paper (Abramowicz et al. 2001).

We should note a few other numerical experiments we have carried out
which provide further insights.  First, as already mentioned, when we
use too large an artificial resistivity in the simulations, such that
the magnetic field reconnects long before it reaches equipartition, we
find that the flow does not make a transition to a
convection-dominated form.  This is not surprising since in this case
there is very little heating from reconnection and therefore there is
not enough entropy production to drive significant convection.
Although it is not fully understood how reconnection works in
astrophysical plasmas (but see the recent work of Lazarian \& Vishniac
1999), it does seem reasonable to assume that significant reconnection
occurs only after the field builds up at least to equipartition
strength (as proposed by Shvartsman 1971).  If this assumption is
valid for real astrophysical flows, then the simulations we have
presented, and our analytical results, should be relevant.

To further investigate the importance of resistive heating, we have
carried out a series of simulations in which we set the artificial
resistivity $\eta$ in equations (3) and (4) to zero.  In these runs,
we find significant magnetic reconnection through numerical
resistivity, but there is no corresponding heating of the gas.  The
flows do not exhibit convection.  As in Models~A--C described in \S3,
when the magnetic field reaches equipartition with the gas pressure, a
central ``core'' region is formed, bounded by a quasi-spherical shock.
However, the core is more compact and the accretion rate is only
slightly reduced with respect to the Bondi rate. The flow pattern in
the core is perturbed with respect to the spherical inflow due to the
effect of the magnetic field, but the perturbations are not as
significant as in Models~A--C, and the velocity streamlines are not as
chaotic.

To understand how important the magnetic field is for the formation of
the convective core, we have carried out three-dimensional
hydrodynamic simulations with finite bulk and/or shear viscosity, and
with two values of $\gamma$: 5/3, 4/3.  We expected that the bulk
viscosity would heat the accreting gas and that this might lead to
efficient convection.  Instead, we find that viscous Bondi flows are
stable for a wide range of values of the bulk and shear viscosity
coefficients.  Combined with the experiments described in the previous
paragraph, the implication is that both the electromotive forces
associated with the magnetic field and the heating effect due to
reconnection are important for the flow to become
convection-dominated; any one by itself is not enough.  Once the CDBF
state has been achieved, it appears to be stable and self-sustaining.

We note recent simulations of three-dimensional rotating MHD accretion
flows by Hawley, Balbus, \& Stone (2001).  These flows exhibit the
magnetorotational instability, as expected.  One might have expected
them also to be convective, at least at large radius, and to have a
radial structure similar to a CDBF (or a CDAF).  However, Hawley et
al. (2001) report that they do not observe convective motions in their
models.  We suspect that this may be because they do not include a
resistive heating term.  (Stone \& Pringle 2001 do include an
artificial resistivity, but their simulations are in two dimensions.)
In the Hawley et al. simulations, the energy release due to
(numerical) reconnection is lost from the system and the total energy
is not conserved.  As noted above, we have simulated spherical
accretion without including an artificial resistivity, and we do not
see convection.  In this sense, the two studies are consistent.

Accretion flows in many astrophysical systems involve magnetized
plasma.  We would like to suggest that any astrophysical system that
has radiatively inefficient spherical accretion will set up a
convection-dominated flow similar to the CDBF solution discussed here,
and will behave very differently from the standard Bondi solution.  In
particular, we suggest that the mass accretion rate will be given by
equation (25), which is very much less than the Bondi accretion rate.
This has potentially important implications.

In a classic paper, Fabian \& Canizares (1988) used the Bondi solution
to estimate the mass accretion rate $\dot M$ onto supermassive black
holes in the nuclei of nearby giant elliptical galaxies and showed
that the observed nuclear luminosities are far below the luminosity
expected if the radiative efficiency is the canonical 10\%.  The
problem has become more severe in recent times with improved
observations of the nucleus of our own Galaxy (Baganoff et al. 2001)
and nuclei of other nearby galaxies (e.g. Di Matteo et al. 2001).  One
solution to the luminosity problem is to assume that the accretion
occurs in a radiatively inefficient mode, e.g. via an ADAF (Narayan,
Yi, \& Mahadevan 1995; Fabian \& Rees 1995; Reynolds et al. 1996).  The
present work suggests an even simpler solution, namely, the mass
accretion rate onto the supermassive black hole may be much less than
the Bondi rate assumed by Fabian \& Canizares (1988).  For conditions
typical of galactic nuclei, say $c_\infty\lesssim10^3 ~{\rm
km\,s^{-1}}$, we expect $R_a\sim10^5R_g$.  Since $R_{in}$ in equation
(25) is likely to be no more than a few tens of $R_g$, we see that
$\dot M$ with a CDBF could be smaller than $\dot M_{Bondi}$ by a large
factor $\sim10^3-10^4$.

Another application is to isolated neutron stars and black holes
accreting from the interstellar medium in the Galaxy.  Treves \& Colpi
(1991) and Blaes \& Madau (1993) used the Bondi accretion rate to
estimate the likely luminosities of accreting neutron stars and
discussed the possibility of detecting them by their EUV and X-ray
emission.  A number of later papers have discussed theoretical
predictions for the emitted spectrum (e.g. Turolla et al. 1994; Zane,
Turolla, \& Treves 1996).  Despite careful searches in the ROSAT
all-sky survey, the predicted large number of sources has not been
found (e.g. Belloni, Zampieri, \& Campana 1997).  As in the case of dim
galactic nuclei, we suggest that the discrepancy is because the
accretion on the neutron stars occurs via a convection-dominated flow,
so that the mass accretion rate is far below the Bondi rate.  There
are similar implications also for accreting stellar-mass black holes
(e.g. Fujita et al. 1998).

Yet another possible application is to supernova explosions.  In
addition to the prompt collapse of a homologous core, current models
of supernovae predict fallback of material over an extended period of
time after the explosion (Chevalier 1989).  This material, which is
ejected with less than the escape speed, flows out radially, turns
around at some (large) radius and collapses back on the compact core.
Some of this material may experience significant magnetic field
amplification and may undergo reconnection and heating.  If so, it is
likely to develop convective motions, resulting in a much reduced rate
of mass fallback.  This deserves further study.

\acknowledgments

We gratefully thank Tom Abel, Marek Abramowicz, Axel Brandenburg and
Eliot Quataert for helpful discussions and comments.  This work was
supported by NSF grant AST 9820686, NASA grant NAG5-10780 and RFBR
grant 00-02-16135.

\appendix

\clearpage

%
%

\clearpage

\begin{deluxetable}{lcc}
\footnotesize \tablecaption{Parameters of the Models \label{tbl-1}}
\tablewidth{0pt} \tablehead{ \colhead{Model~~~~~~} &
\colhead{~~~~~~$\eta_0$\tablenotemark{a}~~~~~~} &
\colhead{~~~~~~$b_0$\tablenotemark{b}~~~~~~} }
\startdata A & 0.5 & 0.3 \nl
B & 0.5 & 0.1 \nl
C & 0.3 & 0.3 \nl
D & 0 & 0 \nl
\enddata
\tablenotetext{a}{$\eta_0$ is a dimensionless artificial resistivity
parameter, defined in equation (8).}
\tablenotetext{b}{$b_0$ characterizes the initial
magnetic field, and is defined in equation (11).}
\end{deluxetable}

\clearpage

\clearpage

\begin{figure}
\caption{Time evolution of the mass accretion rate, normalized to the
Bondi rate, in Model~A (thin line), Model~B (dashed line) and Model~D
(thick line).  Note the suppression of the accretion rate in Models~A
and B, both of which have magnetic fields.  Model~D corresponds to a
pure hydrodynamic flow.
\label{fig2}}
\end{figure}

\begin{figure}
\caption{Magnetic lines in Model~A at time $t=0.5$ (measured in units
of the free-fall time from the edge of the grid).  The cross-section
shown corresponds to the $xz$ plane, and the axes are labeled in
units of the gravitational radius.  Initially, at $t=0$, the magnetic
field is uniform and vertical.  With time, the frozen-in field lines
are deformed under the action of the spherically converging accretion
flow.  The black hole is at the origin.
\label{fig3}}
\end{figure}

\begin{figure}
\caption{Density distribution in the $xz$ plane in Model~A at time
$t=7.88$.  The black hole is located at the origin.  Matter is
concentrated towards the equatorial plane (the horizontal axis) and
towards the black hole.  Note the density inhomogenities which have
been produced by the motion of convective blobs.  There is a shock at
around $80R_g$, where the supersonically infalling gas from the outer
boundary meets the convective core.  The two polar funnels are filled
with low density matter.
\label{fig4}}
\end{figure}

\begin{figure}
\caption{Magnetic field lines in the $xz$ plane in Model~A at time
$t=7.88$.  Except for the polar regions, the magnetic field elsewhere
has a highly tangled morphology.  This is the result of convection.
\label{fig5}}
\end{figure}

\begin{figure}
\caption{Velocity streamlines in the $xz$ plane in Model~A at time
$t=7.88$.  A complicated pattern is seen, with numerous vortices and
eddies.  This is the result of convection.
\label{fig6}}
\end{figure}

\end{document}